\documentclass{ws-ijgmmp-web}
\usepackage{bm}
\usepackage{color}

\newtheorem{rmk}{Remark}
\newtheorem{ex}{Exercise}

\newcommand{\Ignore}[1]{}

\newcommand{\ket}[1]{\left\vert #1\right\rangle}
\newcommand{\bra}[1]{\left\langle #1\right\vert}
\newcommand{\braket}[2]{\langle#1\vert #2\rangle}

\def\e{\mathrm{e}}
\def\ii{\mathrm{i}}
\def\d{\mathrm{d}}

\def\H{\mathcal{H}}

\def\I{\bm{I}}
\def\cU{{\mathcal{U}}}

\begin{document}

\markboth{S. Di Martino, P. Facchi}
{Quantum systems with time-dependent boundaries}
\catchline{}{}{}{}{}
%
%%%%%%%%%%%%%%%%%%%%%%%%%%%%%%%%%%%%%%%%%%%%%%%%%%%%%%%%%%%%%%%%%%%%

\title{QUANTUM SYSTEMS WITH TIME-DEPENDENT BOUNDARIES}

\author{SARA DI MARTINO}
\address{Dipartimento di Matematica, Universit\`a di Bari, Via Orabona 4\\ Bari, I-70125, Italy}

\author{PAOLO FACCHI}
\address{Dipartimento di Fisica and MECENAS, Universit\`a di Bari, Via Amendola 173\\ Bari,  I-70126, Italy\\ INFN, Sezione di Bari, Via Amendola 173\\ Bari, I-70126, Italy}
%\email{paolo.facchi@ba.infn.it}

\maketitle

%\begin{history}
%\received{(Day Month Year)}
%\revised{(Day Month Year)}
%\end{history}

\begin{abstract}
We present here a set of lecture notes on  quantum systems with time-dependent boundaries.
In particular, we analyze the  dynamics of a non-relativistic particle in a bounded domain of physical space, when the boundaries are moving or changing. In all cases, unitarity is preserved and the change of boundaries does not introduce any decoherence in the system. \end{abstract}


\section{Introduction} 

We present here the notes of three lectures given by one of us at the International Workshop on Mathematical Structures in Quantum Physics, held in February 2014 in Bangalore at the Centre for High Energy Physics, Indian Institute of Science.
The course considers some aspects of quantum systems with time-dependent boundaries, a very active area both from the mathematical point of view, see for instance the works of Yajima~\cite{Yajima87,Yajima96}, Dell'Antonio \textit{et al}~\cite{Dell'Antonio} and Posilicano \textit{et al}~\cite{Posilicano07,Posilicano11},  and from a physical perspective. Notable applications arise in different fields ranging from atoms in cavities \cite{HarocheRaimond1,HarocheRaimond2} to ions and atoms in magnetic traps \cite{atoms_ions}, to superconducting quantum interference devices (SQUID) \cite{reviewSQUID}, to the dynamical Casimir effect in  microwave cavities~\cite{photons}. Moreover, the role and importance of boundary conditions at a fundamental level have been stressed in an interesting series of articles, see~\cite{Bala,Wilczek} and references therein, where varying boundary conditions are viewed as a model of space-time topology change. 

These notes will be organized as follows. In the first lecture, in Section~\ref{sec-interval}, we will study the problem of a quantum particle moving in a one-dimensional box. In particular we will introduce and describe the problem in a pedagogical way focussing on the mathematical aspects of the physical model. We will try to give all the mathematical tools needed to a deep understanding of the physics of the problem. Moreover, at the end of the section, we will follow Berry~\cite{berry} and show how, quite surprisingly, even an easy spectral problem like this can lead to intricate and rich dynamics. In the following lectures, we will look at the dynamical evolution with time-dependent boundaries. In particular, in the second lecture, in Section~\ref{sec-MW}, we will discuss the case in which the walls of the bounding box are moving~\cite{MW}, while in the last lecture, in Section~\ref{sec-change}, we will look at the problem of the change of boundary conditions with time~\cite{AFMP}. Finally, in Section~\ref{sec-conc} we will draw some conclusions. 

Interspersed between the lectures we offer a selection of  exercises ranging in difficulty. They are useful to focus the attention on specific points and to test the learning of the reader, therefore we kindly recommend not to skip them.  

\section{Lecture 1: A particle in a one-dimensional box}
\label{sec-interval}

\begin{figure}[!ht]
   \centering
   \includegraphics[scale=0.50]{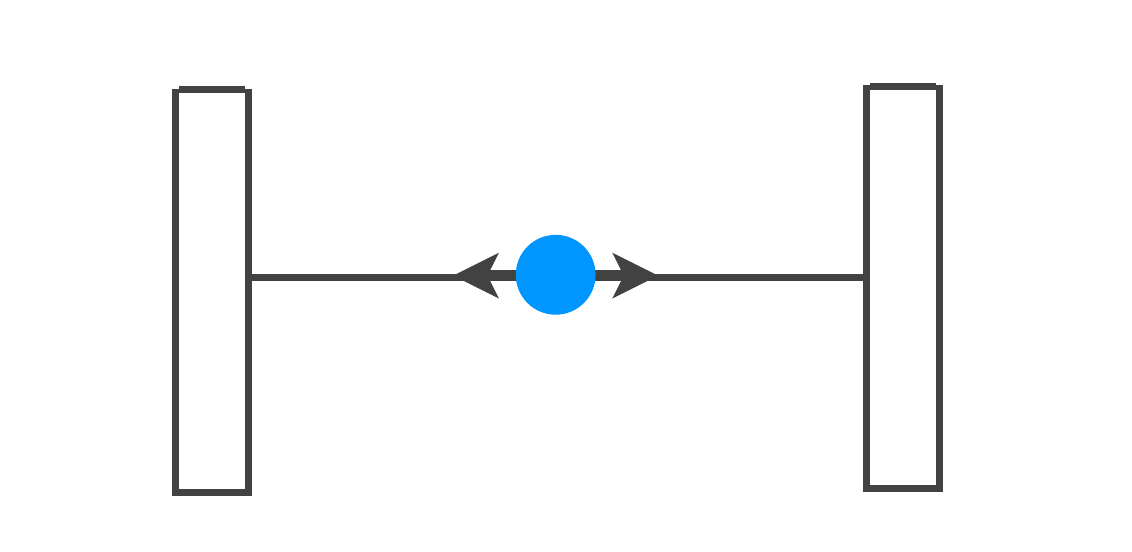}
   \caption{A particle moving in a one-dimensional box.}
   \label{fig:box}
 \end{figure}

One of the very first problems involving boundary conditions in quantum mechanics is the study of a non-relativistic particle in a one-dimensional box. Indeed, it is one of the exercises sometimes  professors give to students just to test their ability. The problem consists in a massive particle moving freely in an infinitely deep well located, say, at $I=[a,b]$ with $b>a$, from which it cannot escape (Fig.~\ref{fig:box}). 

We will see that even if the problem is apparently fairly simple, it leads to very interesting results that stress the difference between classical and quantum mechanics. In fact, the very presence of boundaries in general tends to enhance the quantum aspects of the system. The reason is that in quantum mechanics the behavior at the boundary is constrained by the very structure of the theory and in particular by unitarity. This is at variance with classical mechanics, where the behavior at the boundary does not derive from fundamental principles, and is usually treated phenomenologically.

From the mathematical point of view the system is described by a Hamiltonian operator associated to the kinetic energy of the particle, namely the Laplacian,
\begin{equation}
\label{eq:ham}
T=\frac{p^2}{2m}=-\frac{\hbar^2}{2m}\frac{\d^2}{\d x^2},
\end{equation}
acting on square integrable functions (remember that $|\psi(x)|^2$ is a probability density!) with square integrable second (distribution) derivative on the interval $I$ of the box, $\mathcal{H}^2(I)=\{\psi\in L^2(I)\ |\ p^2\psi\in L^2(I)\}$. In the mathematical literature $\mathcal{H}^2(I)$ is known as the second Sobolev space; it is nothing but the maximal domain of definition of the kinetic energy operator $T$. 

Equation~(\ref{eq:ham}) describes the action of $T$ in the bulk of the box. In order to get a well-defined dynamics, Eq.~(\ref{eq:ham}) should be equipped with suitable \emph{boundary conditions}, which specify the behavior of the particle at the walls. Namely, the wave functions $\psi$ to which we are allowed to apply $T$ must belong to the domain
\begin{equation}
D(T)= \{\psi \in \mathcal{H}^2(I), \text{with suitable boundary conditions}\}.
\end{equation}
The question that here arises naturally is how to choose these boundary conditions. This marks the difference between classical and quantum mechanics: in quantum mechanics the behavior at the boundary, encoded in the domain $D(T)$, derives from basic principles. Indeed, in  quantum mechanics the conservation of probability leads to the  necessity for the evolution to be represented by unitary operators. But unitarity of the evolution is equivalent to  self-adjointness of its generator, as stated in  Stone's theorem \cite{Kato}:
\begin{theorem}{(Stone)}
Let $(U_t)_{t\in\mathbb{R}}$ be a strongly continuous one-parameter group of unitaries on a Hilbert space $\H$. Then there is a unique self-adjoint operator $H$ on $\H$, called the generator of the group, such that $U_t=\e^{-\ii H t/\hbar}$ for all $t$. 
\end{theorem} 
According to  Stone's theorem the operator $T$, that generates the dynamics of the particle, $U_t = \e^{-\ii T t/\hbar}$,  must be self-adjoint, i.e.\ \begin{equation}
T=T^\dagger \qquad \text{and}\quad  D(T)=D(T^\dagger).
\end{equation} 
In order to find the domain $D(T)$ let us consider $\phi,\psi\in \mathcal{H}^2([a,b])$; then a double integration by part gives (prove it!):
\begin{eqnarray}
\label{eq:prod-scal}
-\braket{\psi}{\phi''}&=&- \int_{a}^{b}\overline{\psi(x)}\phi''(x)\, \d x = -\braket{\psi''}{\phi}+ \Lambda(\psi,\phi),
\\
\Lambda(\psi,\phi)&=&\overline{\psi'(b)}\phi(b)-\overline{\psi'(a)}\phi(a)-\overline{\psi(b)}\phi'(b)+\overline{\psi(a)}\phi'(a).
\label{eq:Lambdadef}
\end{eqnarray}
The self-adjointness requirement amounts to finding a maximal subspace $D(T)$ of $\mathcal{H}^2([a,b])$ on which the \emph{boundary form} $\Lambda$ identically vanishes, namely  $\Lambda(\psi,\phi)=0$ for all $\phi, \psi \in D(T)$. 

The more common boundary conditions you can encounter in the literature are:
\begin{enumerate}
\item {\em Dirichlet boundary conditions}:
\begin{equation}
\label{eq:dir}
   \psi(a)=\psi(b)=0;
\end{equation}
\item {\em Neumann boundary conditions}, involving the derivative of the function:
\begin{equation}
\label{eq:neu}
   \psi'(a)=\psi'(b)=0.
\end{equation}
(Of course, the above conditions must be satisfied by \emph{both} $\psi$ and $\phi$ in the boundary form $\Lambda(\psi,\phi)$!)
\end{enumerate}
\begin{ex}
Check that the Dirichlet and the Neumann boundary conditions make $\Lambda$ identically vanish, and thus, modulo a check of  maximality, yield a good domain $D(T)$ of self-adjointness for the kinetic energy operator $T$.
\end{ex}
There is a one-parameter family of boundary conditions that includes Dirichlet and Neumann as limit cases: \emph{Robin's boundary conditions}, that connect the function with its derivative at the boundary:
\begin{equation}
\label{eq:rob}
   \psi'(a)=-\frac{1}{l_0}\tan \frac{\alpha}{2}\, \psi(a), \qquad
\psi'(b)=\frac{1}{l_0}\tan \frac{\alpha}{2}\, \psi(b),
\end{equation}
where $\alpha\in(-\pi,\pi]$, and $l_0$ is a reference length.
From~(\ref{eq:rob}) we recover Dirichlet and Neumann in the limits $\alpha\to\pi$ and $\alpha\to0$, respectively.

Notice that the choice of a domain of self-adjointness $D(T)$ is not just a mathematical nuisance. Different domains give rise to different physics!
Indeed, the  choice of boundary conditions strongly depends on the physical behavior of the walls, as we will now show.

Let us concentrate on one wall, e.g.\ the left one at $a=0$, and look at the scattering process of a plane wave arriving from the left with momentum $-\hbar k <0$. After reflection by the wall a plane wave with momentum $\hbar k$ will arise.
 \begin{figure}[!ht]
   \centering
   \includegraphics[scale=0.50]{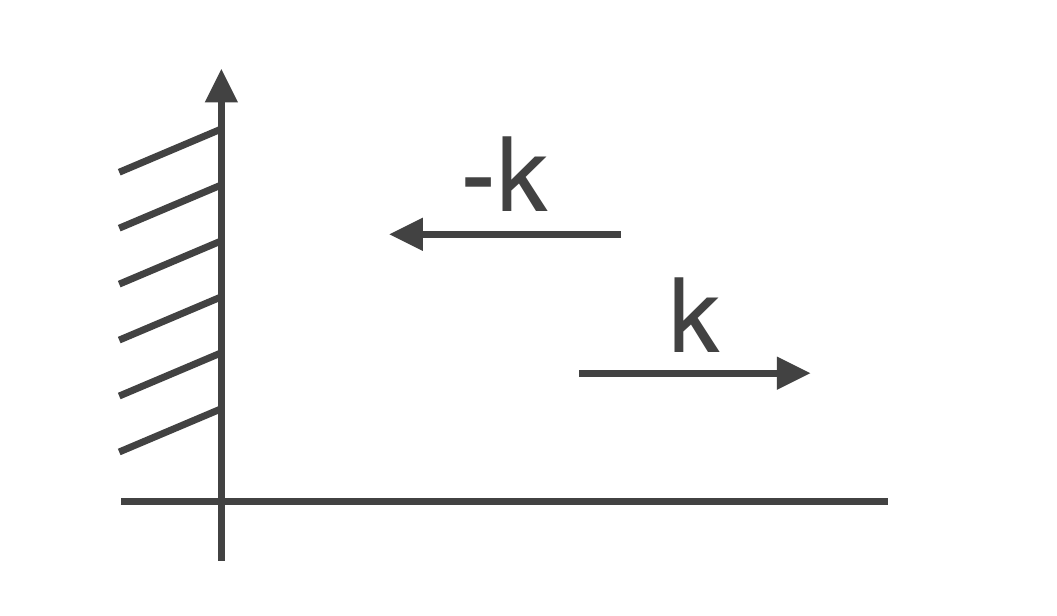}
   \caption{Reflection at a wall}
   \label{fig:walls}
 \end{figure}
Therefore, the wave function will have the form
\begin{equation}
\psi(x)\sim \e^{-\ii kx}+r\,\e^{\ii kx}, 
\end{equation}
where $r$ is a reflection coefficient. Since we are dealing with an impenetrable box, and we assume that the particle does not stick at the wall, we are forced to require that $|r|^2=1$, so that the only freedom left is in choosing a phase, namely $r=\e^{\ii \beta}$. The Dirichlet boundary condition, $\psi(0)=0$, corresponds to a phase $\beta=\pi$ for all $k$, while  Neumann corresponds to the choice $\beta=0$ for all $k$. Notice that in general the phase $\beta(k)$ has a nontrivial dependence on the wavenumber $k$. The following exercise shows that for a generic $\alpha$ in~(\ref{eq:rob})  the behavior of the walls depends on the momentum $\hbar k$ of the impinging particle  through a nontrivial phase $\beta(k)$.
\begin{ex}
Prove that Robin's boundary condition $\psi'(0)=-\frac{1}{l_0}\tan \frac{\alpha}{2}\, \psi(0)$ corresponds to a momentum dependence of the phase shift $\beta(k)$ given by
\begin{equation}
\tan \frac{\beta(k)}{2} = \frac{1}{k\, l_0} \tan \frac{\alpha}{2}.
\end{equation}
Recover, as a limit, Dirichlet and Neumann, and show that they are the only phases independent of $k$.
\end{ex}

The general theory of self-adjoint extensions and boundary conditions \cite{AIM} asserts that
all possible boundary conditions that make $T$ self-adjoint can be parametrized by $2\times 2$ unitary matrices $U$ in the following way:
\begin{equation}
\ii (\I +U) \Psi' = (\I-U) \Psi,  \qquad U\in\cU(2),
\label{eq:bc2}
\end{equation}
where
\begin{equation}
\Psi := \left(\begin{array}{r} \psi(a)  \\
\psi(b) 
\end{array}\right), \qquad \Psi' := l_0\left(\begin{array}{r} 
- \psi'(a) \\
\psi'(b)\end{array}\right),
\label{eq:bc21}
\end{equation}
and $\I$ is the identity matrix.
\begin{ex}
Check that if $\psi$ and $\phi$ satisfy~(\ref{eq:bc2})-(\ref{eq:bc21}) for the same unitary $U$, then the boundary form~(\ref{eq:Lambdadef}) vanishes, $\Lambda(\psi,\phi)=0$.
\end{ex}

Robin's boundary conditions~(\ref{eq:rob}) correspond to (prove it!)
\begin{equation}
U=\e^{-\ii\alpha} \I
=\left(\begin{array}{l l} 
\e^{-\ii\alpha} &0  \\
0&\e^{-\ii\alpha} 
\end{array}\right),
\end{equation} 
and in particular the Dirichlet boundary conditions correspond to $U=-\I$, while the Neumann ones are given by $U=\I$.
More generally, by taking
\begin{equation}
U=\left(\begin{array}{l l} 
\e^{-\ii\alpha_1} &0  \\
0&\e^{-\ii\alpha_2} 
\end{array}\right),
\label{eq:local}
\end{equation} 
with $\alpha_1, \alpha_2 \in (-\pi,\pi]$ we can describe different boundary conditions at the two endpoints, e.g.\ Dirichlet/Neumann ($\alpha_1=\pi$, $\alpha_2=0$), and include all cases considered above, when $\alpha_1=\alpha_2$.
 But, of  course the unitaries~(\ref{eq:local}) do not exhaust all allowed boundary conditions, since they form only a two-dimensional submanifold $\mathcal{U}(1)\times \mathcal{U}(1)$ of the total four-dimensional manifold $\mathcal{U}(2)$. 
 Thus, where do  all  missing boundary conditions come from? 
 
 Up to now we have considered only \emph{local} boundary conditions, somewhat deceived by the physics depicted in Fig.~\ref{fig:walls}: the two walls did not talk each other. Indeed, the unitary matrices in~(\ref{eq:local}) are diagonal and do not mix the boundary values at $a$ with those at $b$; we missed all nondiagonal unitaries, describing \emph{nonlocal} boundary conditions. Mathematics tells us that unitarity allows also for them. So what physics do they describe, if any?
 \begin{figure}[!ht]
   \centering
   \includegraphics[scale=0.25]{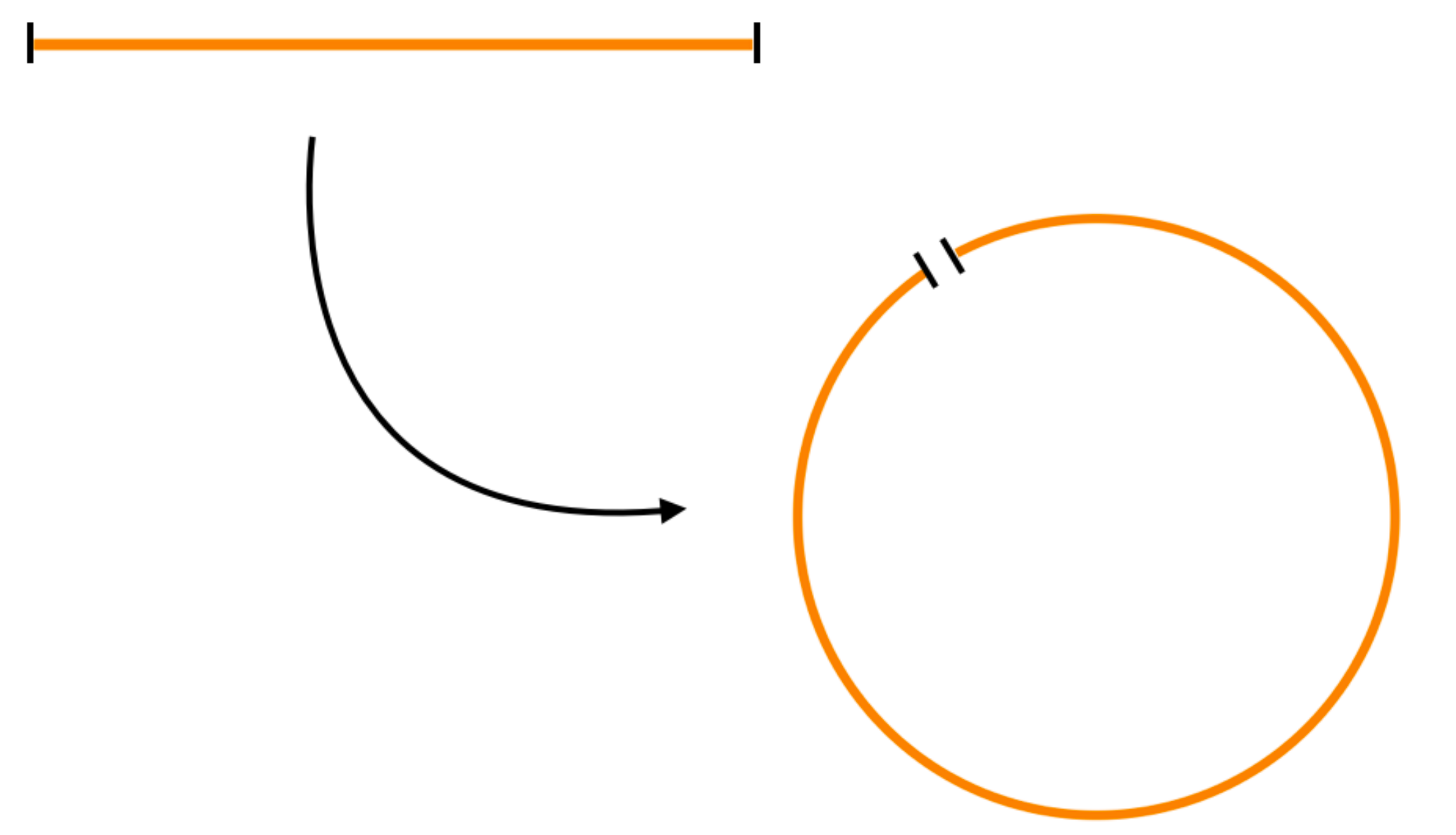}
   \caption{Change of topology: from an interval to a circle}
   \label{fig:bend}
 \end{figure}
 
 In fact, nondiagonal unitaries are describing a physical situation which is different from that of a box with two walls far apart. In order to interact, the two ends of the interval should come close, as in Fig.~\ref{fig:bend}, so that the interval should be bent and the two walls should become the two sides of a junction. In other words, by changing from a diagonal $U$ to a nondiagonal $U$ we are assisting at a  change of topology: from an interval to a circle.
 
Therefore, the geometry that is able to support all possible boundary conditions in $\mathcal{U}(2)$ is that of a ring with a junction. If the junction is impermeable, i.e. there is total reflection at the walls, we are back to the interval, otherwise there is nonzero transmission across the junction, from one wall to the other. An interesting example is given by the matrix
\begin{equation}
U= \left(\begin{array}{ll}0 & \e^{-\ii\alpha}  \\ \e^{\ii\alpha} &\ 0 \end{array}\right) = 
\cos \alpha\, \sigma_x + \sin\alpha\, \sigma_y,
\label{eq:Uperiodic}
\end{equation}
which describes
pseudo-periodic boundary conditions (prove it!):
\begin{equation}
\psi(b)=\e^{\ii \alpha}\, \psi(a), \qquad \psi'(b)=\e^{\ii \alpha}\, \psi'(a).
\label{eq:phiperiodic}
\end{equation}
By passing through the junction, the wave function acquires a phase $\alpha$. If $\alpha=0$ we get the famous periodic boundary conditions, and the geometry is that of a circle; if $\alpha=\pi$ we get antiperiodic boundary conditions. The phase $\alpha$ encodes the properties of the junction, for example the material it is made of, or its width.

In general, if the unitary $U$ has no $-1$ eigenvalues, the wave function $\psi$ can assume any value at the endpoints of the interval. Only the boundary values of its derivative are constrained in some way. On the other hand, one eigenvalue equal to $-1$ corresponds to one constraint on the values of $\psi$ at the ends, as for example in the first equation in~(\ref{eq:phiperiodic}). Finally, two $-1$ eigenvalues, i.e.\ $U=-\I$, correspond to  two constraints on the boundary  values of the wave function, i.e.\ Dirichlet at both ends.
\begin{ex}
Prove that the unitary~(\ref{eq:Uperiodic}) has always an eigenvalue equal to $-1$. Find its corresponding eigenvector $\xi$ and show that the first equation in~(\ref{eq:phiperiodic}) is nothing but an orthogonality condition $\braket{\xi}{\Psi}=0$. (In Lecture 3 we come back to the geometrical meaning of this condition.)
\end{ex}

\subsection{Hard walls}
\begin{figure}[!ht]
   \centering
   \includegraphics[scale=0.50]{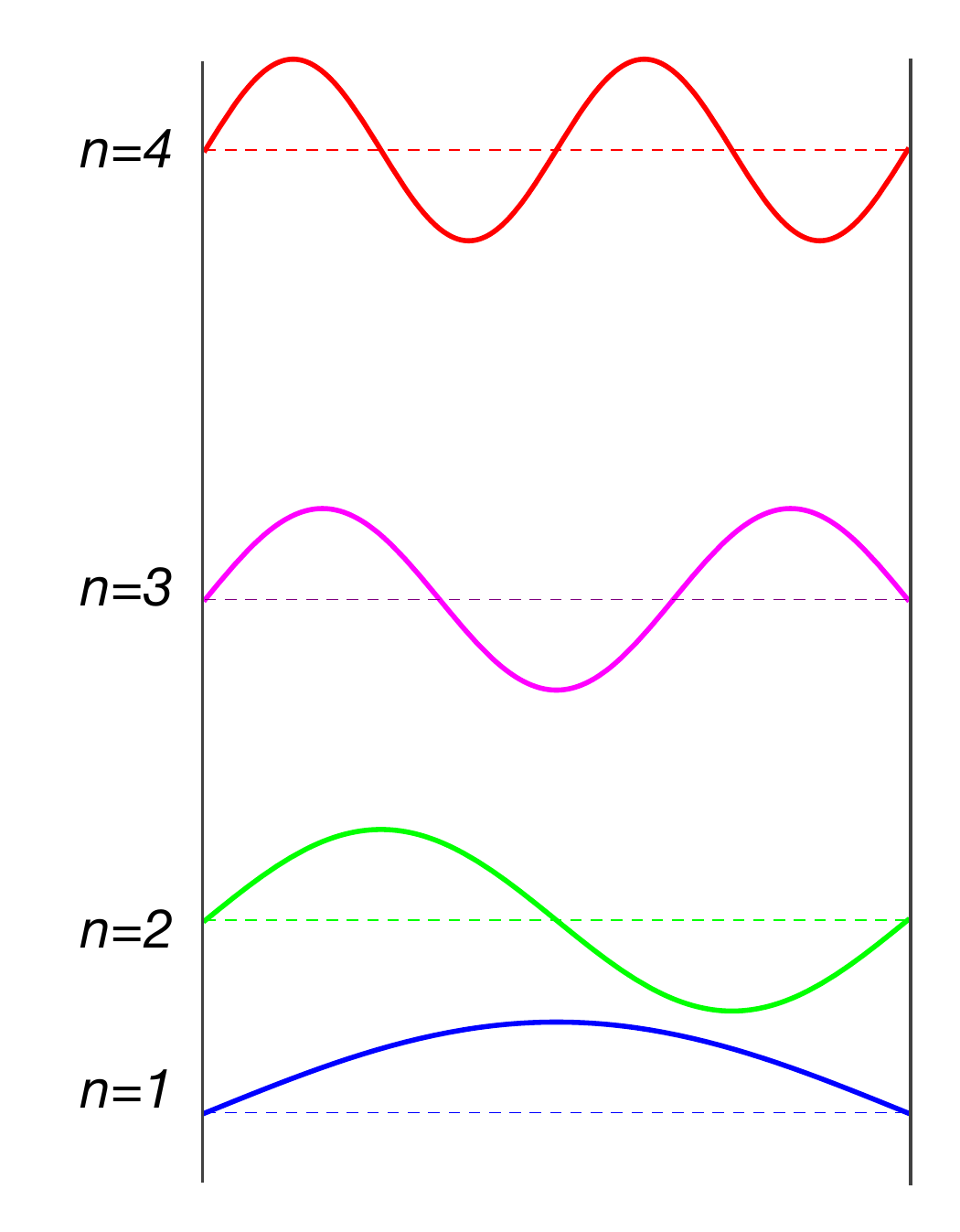}
   \caption{The eigenfunctions (solid lines) of the operator $T$ in the box at different levels of energy (dashed lines).}
   \label{fig:eigen}
 \end{figure}

Let us concentrate now on the textbook case of a particle confined in an infinitely deep well, Fig.~\ref{fig:box}. The appropriate boundary conditions are Dirichlet's: $\psi(a)=\psi(b)=0$.
The dynamics of the particle is described by the  Schr\"{o}dinger equation
\begin{equation}
\ii\hbar \frac{\partial \psi(x,t)}{\partial t}= -\frac{\hbar^2}{2m} \frac{\partial^2 \psi(x,t)}{\partial x^2}.
\label{eq:schr_eq}
\end{equation}
A separation of  variables, $\psi(x,t)=u(x) \exp(-\ii E t/\hbar)$,  reduces the problem to  the solution of the spatial part  of the differential equation, that means to find  eigenvectors and  eigenvalues of the operator $T$:  
\begin{equation}
-\frac{\hbar^2}{2m}u''(x)=E\, u(x).
\end{equation} 
The general solution is 
\begin{equation}
u(x)=c_1\, \e^{+\ii k x}+c_2\,\e^{-\ii k x},
\end{equation} 
with $k = \sqrt{2 m E}/\hbar$ (in principle $k$ can be imaginary, but see below), and 
$c_1$ and $c_2$ are arbitrary constants that can be fixed (up to a common phase) by imposing  the Dirichlet boundary conditions,
\begin{eqnarray}
u(a)&=&c_1\, \e^{\ii k a}+c_2\, \e^{-\ii k a}=0\\
u(b)&=&c_1\, \e^{\ii k b}+c_2\,\e^{-\ii k b}=0,
\end{eqnarray}
and  normalization
\begin{equation}
\braket{u}{u}=1.
\end{equation}
\begin{ex}
Prove that the normalized eigenfunctions of $T$, with the Dirichlet boundary conditions, are
\begin{equation}
\label{eq:eigen}
u_n(x)=\sqrt{\frac{2}{l}} \sin \left(\frac{n \pi}{l} (x-a)\right),
\end{equation}
where $l=b-a$, and that the eigenvalues, giving the permitted energy levels are
\begin{equation}
\label{eq:energy}
E_n=\frac{\hbar^2}{2m}\frac{n^2\pi^2}{l^2},
\end{equation}
for $n=1,2,\dots$. See Fig.~\ref{fig:eigen}
\end{ex}
\begin{rmk}
Eq.~(\ref{eq:energy}) tells us two important things on the bound states of the particle:
\begin{enumerate}
\item the energies are quantized;
\item the energy is always strictly positive. 
\end{enumerate} 
\end{rmk}
The strictly positivity of the energy is a property of the Dirichlet boundary conditions. Indeed, with the Neumann ones the energy of the ground state is 0, and, more surprisingly, it is even \emph{negative} with Robin's boundary conditions!
\begin{ex}
Prove that the ground state of $T$, with the Neumann boundary conditions is
\begin{equation}
v_0(x)=\sqrt{\frac{1}{l}},
\label{eq:v0}
\end{equation}
and has zero energy, $E_0=0$. Then, look at the eigenvalue problem with Robin's boundary conditions~(\ref{eq:rob}).
\end{ex}

\subsection{Fractals in a box}
The textbook exercise  of the quantum particle in a box inevitably ends with the evaluation of the eigenvalues~(\ref{eq:energy}) and the eigenfunctions~(\ref{eq:eigen}). The result is so simple and intelligible that we all felt a profound satisfaction when we derived it in our first course of quantum mechanics. The simplicity of the spectrum is deceptive and leads us to think that we fully understand the physical problem. In particular, we are convinced that the dynamics, which is the solution to the Schr\"odinger equation~(\ref{eq:schr_eq}), must surely be as much simple. In fact, this belief is false, as showed by Berry~\cite{berry}: the dynamics is instead very intricate.

Let us assume  at time $t=0$ that $\psi(x,0)=v_0(x)$, with $v_0$ given by~(\ref{eq:v0}). This is the simplest conceivable initial condition, corresponding to a flat probability in the box $[a,b]$, with $l=b-a$.
\begin{figure}[!ht]
   \centering
   \includegraphics[scale=0.40]{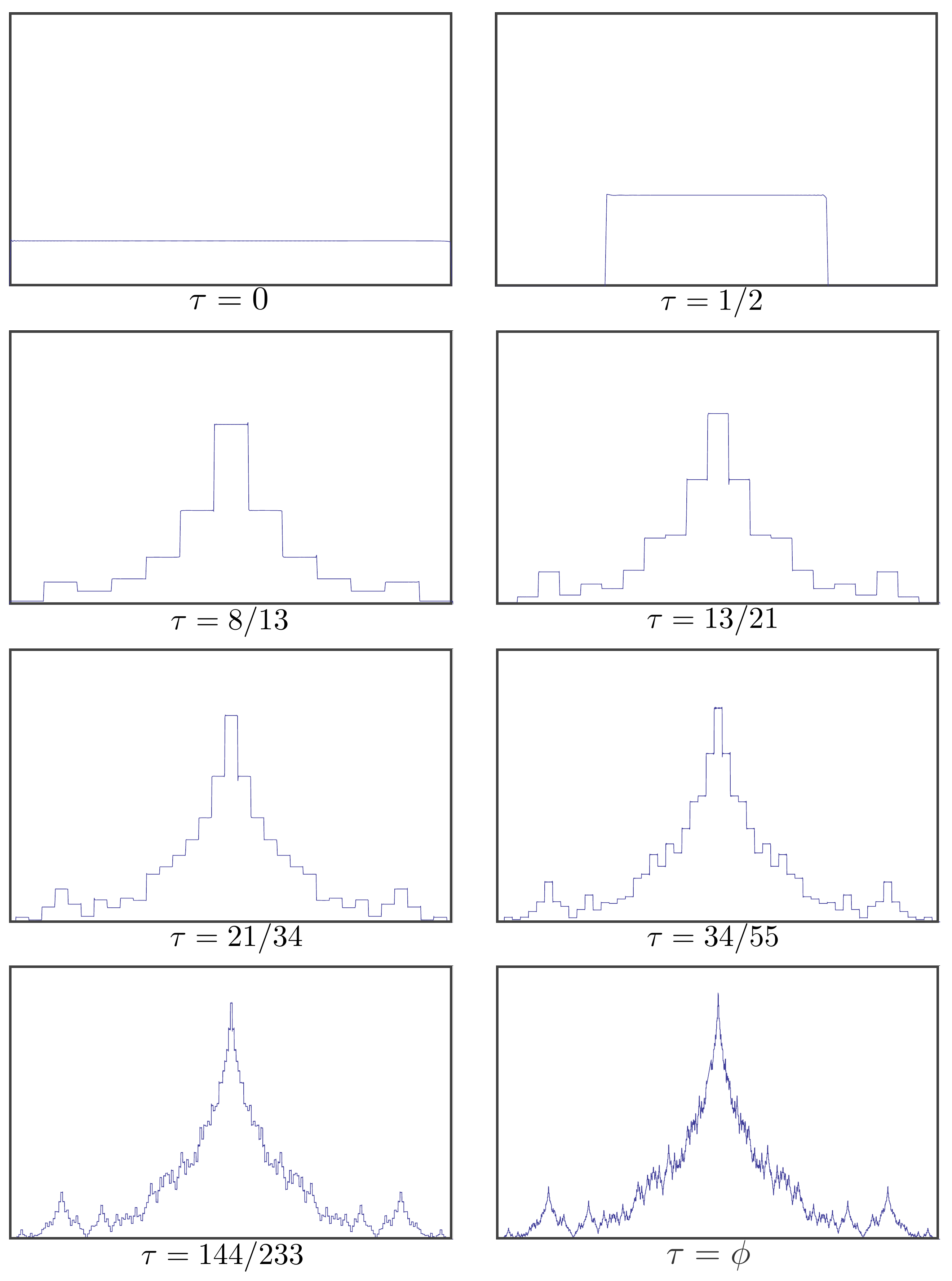}
   \caption{Graphs of $|\vartheta(\xi,\tau)|^2$ vs $\xi$ at different rational times $\tau$ along the Fibonacci sequence tending to the golden mean, $\phi=(1+\sqrt{5})/2$. See the emergence of a fractal structure.}
   \label{fig:Fractals}
 \end{figure}
We are interested in the time evolution of this initial wave function. Its $L^2$-expansion in terms of the eigenfunctions~(\ref{eq:eigen}) of $T$ reads
\begin{equation}
v_0(x)=\sum_{n=1}^\infty c_n u_n(x),
\end{equation}
where $c_n=\braket{u_n}{v_0}$. 
\begin{ex}
Show that 
\begin{equation}
c_n=\frac{\sqrt{2}}{n\pi} \left[1-(-1)^n\right],
\label{eq:cn}
\end{equation} 
and, in particular, $c_{2n}=0$, for all $n=1,2,\dots$.
\end{ex}
In the same way the quantum evolution, described by the action of the unitary operator $U(t)=\e^{-iTt/\hbar}$, can be written as an $L^2$-convergent series
\begin{equation}
\psi(x,t)=(\e^{-\ii T t/\hbar}v_0)(x)=\sum_{n=1}^{+\infty}c_n\ \e^{-\ii E_n t/\hbar}\ u_n(x).
\end{equation}
We can now use the explicit expressions~(\ref{eq:eigen})-(\ref{eq:energy}) and obtain
\begin{equation}
\psi(x,t)=\sqrt{\frac{2}{l}}\  \sum_{n=1}^{+\infty}c_n \sin \left(\frac{n \pi}{l} (x-a)\right)\exp\left(-\frac{\ii \hbar}{2m}\frac{n^2\pi^2 t}{l^2}\right) .
\end{equation}
In terms of the dimensionless variables $\xi = l^{-1} (x- (a+b)/2) \in [-1/2, 1/2]$, and $\tau= 2 \pi\, t\, \hbar /m l^2 \in\mathbb{R}$, it reads
\begin{equation}
\vartheta(\xi,\tau)=\sqrt{\frac{2}{l}}\  \sum_{k=0}^{+\infty}c_{2k+1} \sin \left[2\pi\left(\xi+\frac{1}{2}\right) \left(k+\frac{1}{2}\right)  \right]\exp\left[-\ii \pi \tau \left(k+\frac{1}{2}\right)^2 \right] .
\end{equation}
By writing the sine as the sum of exponentials and by making use of the expression~(\ref{eq:cn}), we finally get
\begin{equation}
\vartheta(\xi,\tau)=\sum_{n=-\infty}^{+\infty} d_n \e^{\ii 2 \pi  \xi \left(n+\frac{1}{2}\right) -\ii \pi \tau \left(n+\frac{1}{2}\right)^2}, \qquad \xi\in \left[-\frac{1}{2}, \frac{1}{2}\right],
\label{eq:finalsum}
\end{equation}
(notice that now the sum runs over all $n\in\mathbb{Z}$). 
\begin{ex}
Derive Equation~(\ref{eq:finalsum}) and show that
\begin{equation}
\label{dn}
d_n= \frac{1}{\pi \sqrt{l}} \frac{(-1)^n}{n+\frac{1}{2}}, \qquad n\in\mathbb{Z}.
\end{equation}
\end{ex}
If we take a closer look at the expression~(\ref{eq:finalsum})  we notice that it is a Fourier series with quadratic phases. This series is the boundary value of a Jacobi theta function~\cite{Abra}, which is defined in the lower complex half plane of $\tau$, and it has a very rich structure investigated at length by mathematicians. For a full immersion in its deep arithmetic properties see the charming ``Tata Lectures on Theta'' by Mumford~\cite{theta}.
 A simple property is its quasi-periodicity in the rescaled time $\tau$ (check it!): \begin{equation}
\vartheta(\xi,\tau+1)= \e^{-\ii \pi/4}\, \vartheta(\xi,\tau).
\end{equation} 
Thus at integer times $\tau$ the wave function comes back (up to a phase) to its initial flat form~(\ref{eq:v0}): these are the \emph{quantum revivals}. More generally, at rational values of $\tau$  the graph of $|\vartheta(\xi,\tau)|^2$ is piecewise constant and there is a partial reconstruction of the initial wave function \cite{berryklein}, see Fig.~\ref{fig:Fractals}.  On the other hand, at irrational times, the wave function is a fractal, with Hausdorff dimension $D_H=3/2$, as shown in the last panel of Fig.~\ref{fig:Fractals}. In fact, $\vartheta(\xi,\tau)$ can be proved to be a fractal function in space \emph{and} time, and  to form a beautifully intricate \emph{quantum carpet}, with different Hausdorff dimensions along different space-time directions \cite{berryschleich}.

 \section{Lecture 2: Moving walls!}
\label{sec-MW}
In this lecture we will try to answer the following question: What happens if the walls of our box start moving? Which equation will describe the quantum dynamics of the bouncing particle?

The classical version of this problem was introduced  by Fermi \cite{Fermi} in 1949, and then investigated by Ulam \cite{Ulam}.
It is convenient to parametrize the confining interval as
\begin{equation}
I_{l,d}= \left[ d -\frac{l}{2}, d+ \frac{l}{2}\right],
\end{equation} 
so that $l>0$ is the width of the box and $d\in\mathbb{R}$  its center. We will suppose that $l$ and $d$ are regular functions of time, $t\mapsto l(t)$ and $t\mapsto d(t)$, with $l(t)>l_0>0$ so that the interval never shrinks to  a point.  From now on we will often omit the dependence on $t$.

As we did in the case of still walls, we analyze the dynamics described by the Schr\"odinger equation
 \begin{equation}
\ii \hbar \frac{\d }{\d t} \psi(t)=\frac{p^2}{2m}\psi(t),
\label{eq:Mov_SchrEq}
\end{equation}
where the domain of $p^2$ is
\begin{equation}
D_{l,d}=\left\{ \psi\in
\mathcal{H}^2\left(I_{l,d}\right), \; \psi\left(d-\frac{l}{2}\right)=\psi\left(d+\frac{l}{2}\right)=0\right\},
\label{eq:Dld}
\end{equation}
(Dirichlet's boundary conditions). Notice that this domain depends on time, so that at different times we  work on different spaces. This means that the time derivative, 
\begin{equation}
\frac{\d }{\d t} \psi(t)=\lim_{\epsilon\to 0} \frac{\psi(t+\epsilon)-\psi(t)}{\epsilon},
\end{equation}
involves the sum of vectors belonging to different Hilbert spaces, since in general $D_{l(t),d(t)}\neq D_{l(t+\epsilon),d(t+\epsilon)}$. Therefore, we need to take more care in the formulation of the problem and in the interpretation of Eq.~(\ref{eq:Mov_SchrEq}).

The correct formulation of the problem can be accomplished by embedding the space of square integrable functions on the interval, $L^2(I_{l,d})$, in the larger Hilbert space $L^2(\mathbb{R})$  on the real line:
\begin{equation}
L^2(\mathbb{R})=L^2(I_{l,d})\oplus L^2(I_{l,d}^c),
\label{eq:L2sum}
\end{equation}
where $X^c=\mathbb{R}\setminus X$ denotes the complement of the set $X$.  Thus, every wave function $\psi\in L^2(\mathbb{R})$ can be written as a sum $\psi=\chi +\phi$, where $\chi\in L^2(I_{l,d})$ and $\phi\in L^2(I_{l,d}^c)$. Following this strategy the Hamiltonian of the system, representing the kinetic energy of the particle in the box, in the containing space reads:
\begin{equation}
H_0(l,d)=\frac{p_{l,d}^2}{2m}= -\frac{\hbar^2}{2 m} \frac{\mathrm{d}^2}{\mathrm{d} x^2}\oplus_{l,d} 0,
\label{Ham}
\end{equation}
equipped with the Dirichlet boundary conditions.
Notice that the choice of the Hamiltonian to be 0 in the component $L^2(I_{l,d}^c)$, i.e.\ outside the box,
is completely arbitrary and immaterial, since we will be interested in the dynamics inside the box, for a particle with initial (and evolved) wave function with $\phi=0$. Moreover, it is worth noticing that even if the Hamiltonian in Eq.~(\ref{eq:Mov_SchrEq}) appeared to be time-independent, the direct sum decomposition in~(\ref{Ham}), in the case of moving walls, makes its time dependence clear. Indeed, the direct sum in Eq.~(\ref{eq:L2sum}) depends on time through the functions $l$ and $d$. In other words, the restriction to the interval somehow concealed the time-dependence of the Hamiltonian, which becomes evident once one enlarges the space. Therefore, the Schr\"odinger equation we are dealing with has a time-dependent Hamiltonian on a time-dependent domain. 

\subsection{Reduce and Conquer}

\begin{figure}[!ht]
   \centering
   \includegraphics[scale=0.45]{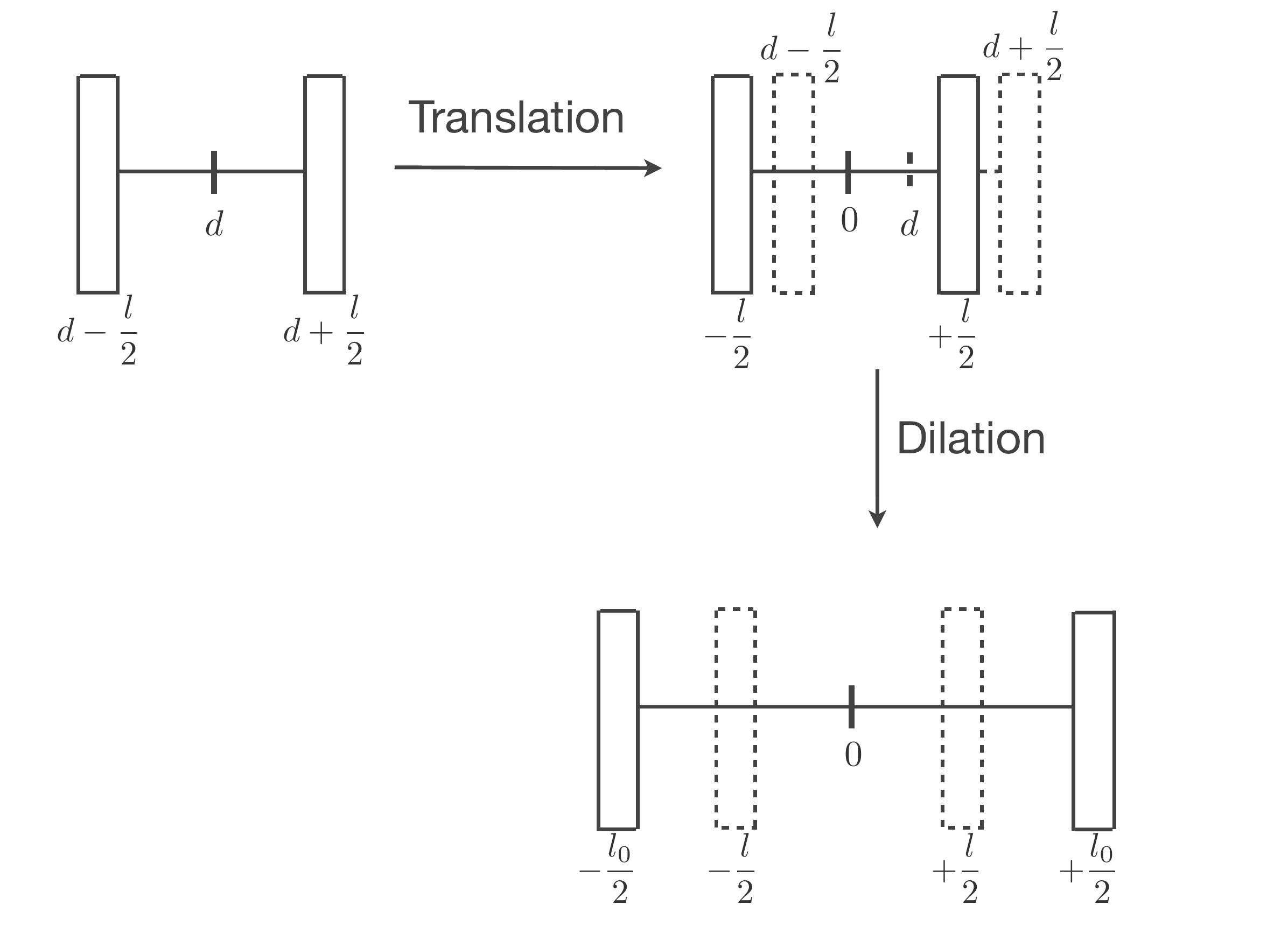}
   \caption{The effect of the transformation on the box can be described as the composition of  a translation and a dilation.}
   \label{fig:transf}
 \end{figure}

The  attack strategy that we will follow 
is to find a unitarily equivalent  problem where the Schr\"odinger operators  act on a \emph{common fixed domain} of self-adjointness.
Let us consider the transformation
\begin{equation}
\qquad U(l,d):L^2(\mathbb{R}) \rightarrow L^2(\mathbb{R}),
\label{eq:Uld}
\end{equation}
that acts as
\begin{equation}
(U(l,d)\psi)(\xi)=\sqrt{\frac{l}{l_0}}\, \psi\left(\frac{l}{l_0} \xi+d\right),
\end{equation}
where $l_0>0$ is a reference length.
\begin{ex}
Prove that the transformation $U(l,d)$ is unitary and
 maps the interval $I_{l,d}$ onto the reference interval
\begin{equation}
I=I_{l_0,0} = \left[-\frac{l_0}{2},\frac{l_0}{2}\right].
\end{equation}
\end{ex}
In fact, $U(l,d)$ can be rewritten as a composition of two transformations  
\begin{equation}
U(l,d) = \mathcal{D}\left(\ln \frac{l}{l_0}\right)^\dag \mathcal{T}(d)^\dag = \mathcal{D}\left(-\ln \frac{l}{l_0}\right) \mathcal{T}(-d),
\label{eq:decomp}
\end{equation}
where
\begin{equation}
(\mathcal{T}(d) \psi) (x) = \psi (x-d), \qquad (\mathcal{D}(s) \psi ) (x) = \e^{-s/2} \psi(\e^{-s} x),
\label{eq:decomp1}
\end{equation}
are two (strongly continuous) one-parameter unitary groups implementing the translations and dilations on $L^2(\mathbb{R})$, see Fig.~\ref{fig:transf}.
\begin{ex}
Prove the decomposition~(\ref{eq:decomp}).
\end{ex}

Now notice that the domain~(\ref{eq:Dld}) is mapped onto the
fixed domain $D=U(l,d) D_{l,d}$
\begin{equation}
 D=\left\{
\phi\in \mathcal{H}^2(I),\; 
 \phi\left(-\frac{l_0}{2}\right)=\phi\left(\frac{l_0}{2}\right)=0\right\}\subset L^2(I),
\label{eq:Dirichlet}
\end{equation}
describing the Dirichlet boundary conditions in a fixed box $I$, and the Hamiltonian~(\ref{Ham}) is mapped into
\begin{equation}
H(l) = U H_0 U^{\dagger}= \left(\frac{l_0}{l}\right)^2
\frac{p^2}{2m} \oplus 0 =-\left(\frac{l_0}{l}\right)^2 \frac{\hbar^2}{2 m}
\frac{\mathrm{d}^2}{\mathrm{d} x^2}\oplus 0.
\label{eq:H(l,d)}
\end{equation}

\begin{ex}
Prove that the unitary transformation $U(l,d)$ yields constant boundary conditions only when the latter do not mix the wave function with  its derivative.
\end{ex}

Henceforth we will denote the wave functions in the frame with moving and fixed walls  by $\psi(x)$ with $x\in I_{l,d}$  and by $\phi(\xi)$ with $\xi\in I$, respectively.
By  deriving the relation 
\begin{equation}
\phi(t)=U(l(t),d(t))\psi(t),
\end{equation}
we get the Schr\"odinger equation in the fixed reference frame from the one in the frame with moving walls:
\begin{eqnarray}
\nonumber
\ii\hbar\frac{\d}{\d t}\phi&=&\ii \hbar\left(U(l,d)\dot{\psi}+\dot{U}(l,d)\psi\right)\\
\nonumber&=&\left(U(l,d)H_0(l,d)+\ii\hbar\dot{U}(l,d)\right)\psi\\
&=&\left(U(l,d)H_0(l,d)U^\dagger(l,d)+\ii\hbar\ \dot{U}(l,d)U^\dagger(l,d)\right)U(l,d)\psi.
\end{eqnarray}
Notice that now the Schr\"odinger operator contains not only the transformed Hamiltonian $H(l)$, but also an additional geometrical term
\begin{equation}
K(l,d)=\ii\hbar \dot{U}(l,d)U^\dagger(l,d).
\label{eq:K(l,d)}
\end{equation}
Let us compute this geometric contribution step by step. First of all the
action of $\dot{U} = \d U(l(t),d(t))/\d t$ on a test function $\psi$:
\begin{eqnarray}
\left(\frac{\d U}{\d t} \psi \right)(\xi)&=&\frac{\d}{\d
t} \left(\sqrt{\frac{l}{l_0}} \, \psi \left(\frac{l}{l_0}\xi+d\right) \right)
\nonumber\\
&=& \frac{\dot{l}}{2\sqrt{l_0 l}}
\psi \left(\frac{l}{l_0}\xi+d\right) +\sqrt{\frac{l}{l_0}}\left(\frac{\dot{l}}{l_0}\xi+\dot{d}\right) \psi' \left(\frac{l}{l_0}\xi+d\right).
\end{eqnarray}
Then, since
\begin{equation}
(U^\dagger(l,d) \phi)(x) =   (\mathcal{T}(d) \mathcal{D}(\ln l/l_0)  \phi)(x) = \sqrt{\frac{l_0}{l}}\,
\phi\left(\frac{l_0}{l}(x-d)\right),
\end{equation}
we have
\begin{equation}
\ii \hbar \frac{\d U}{\d t} U^\dagger \phi (\xi) =  \ii
\frac{\hbar}{2}\, \frac{\dot{l}}{l}\, \phi (\xi) + \ii \hbar  \left( \frac{\dot{l}}{l}\xi+\frac{l_0 }{l} \dot{d}\right) \phi'
(\xi),
\end{equation}
that is
\begin{equation}
\ii \hbar \frac{\d U}{\d t} U^\dagger  =  - \frac{\dot{l}}{l} \left(x  p- \ii
\frac{\hbar}{2}\right)  - \frac{l_0 }{l} \dot{d}\,  p,
\end{equation}
with $x$ and $p$  the  position and momentum operators. Thus, the geometric generator of the unitary transformation reads
\begin{equation}
K(l,d)= \ii \hbar\, \frac{\d U}{\d t} U^\dagger= - \frac{\dot{l}}{l} x \circ p-\frac{l_0 }{l} \dot{d}\, p,
\label{eq:transport}
\end{equation}
where  $A\circ B = (AB+BA)/2$ is the symmetrized (Jordan) product
of the operators $A$ and $B$, and the canonical commutation relation $[x, p]=\ii\hbar$ has been used.
\begin{ex}
Show that the generator of the translation group $\mathcal{T}(d)$ in~(\ref{eq:decomp1}) is the momentum operator $p$, while the generator of the dilation group $\mathcal{D}(s)$ is the virial operator $x\circ p$. As a consequence, by using the decomposition~(\ref{eq:decomp}) show that 
\begin{equation}
U(l,d)= \exp\left(\frac{\ii}{\hbar} \ln \left(\frac{l}{l_0 }\right) x \circ p \right)\, \exp\left(\frac{\ii}{\hbar} d\, p \right),
\end{equation}
which yields~(\ref{eq:transport}).
\end{ex}
Summing up, we finally come to the Schr\"odinger equation in the reference frame with fixed walls 
\begin{equation}\label{eq:StaticH_SchrEq}
\ii \hbar \frac{\d }{\d t}\phi=
\Big(H(l)+K(l,d) \Big)\phi
=\left(\frac{1}{l^2}\frac{p^2}{2m
}-\frac{\dot{l}}{l} x \circ p-\frac{l_0}{l} \dot{d}\, p\right)\phi\,.
\end{equation}

Let us now make some  considerations about this equation and its solution. We need a definition and a theorem \cite{Kato}.

Let $T$ and $A$ be two operators on a Hilbert space $\mathcal{H}$, such that $D(T)\subset D(A)$. $A$ is said to be \emph{relatively bounded} with respect to $T$ or simply \emph{$T$-bounded}, if there exist two non-negative constants $a$ and $b$ such that:
\begin{equation}
\|A \psi \| \leq a \|\psi\|+b \|T \psi\|, \qquad \forall\ \psi\in D(T).
\end{equation}  
Moreover, the infimum of the possible values of $b$  is called the {\it $T$-bound} of $A$.
\begin{theorem}{(Kato-Rellich)} Let $T$ be self-adjoint and bounded from below. If $A$ is symmetric and $T$-bounded with $T$-bound smaller than $1$, then $T+A$ is self-adjoint and bounded from below.
\end{theorem} 

\begin{ex}
Show that $K(l,d)$ is relatively bounded with respect to $H(l)$ with $0$ relative bound.
\end{ex}
 
Applying the Kato-Rellich theorem it is easy to prove that the total Hamiltonian $H(l) + K(l,d)$ with domain $D(H(l)+K(l,d))= D(H(l))=D$ is self-adjoint. Therefore, the Schr\"odinger equation in~(\ref{eq:StaticH_SchrEq}) is well defined for any initial condition $\phi(0)\in D$, and yields a unique unitary propagator. The proof is a corollary of Theorem~X.70 in~\cite{Reed}, since $t\mapsto H(l(t)) + K(l(t),d(t))$ is a one-parameter family of Schr\"{o}dinger operators on a common domain of self-adjointness $D$ for any pair of differentiable functions $d(t)$ and $l(t)$, with $l(t) > l_0$, for some $l_0>0$.

\begin{ex}
Prove that the energy rate equation of the particle is given by
\begin{eqnarray}
\frac{\d}{\d t}\braket{\phi(t)}{H(l(t)) \phi(t)} &=& -\frac{\hbar^2}{2 m} \left(\frac{l_0}{l(t)}\right)^3 \left[\left( \frac{\dot{l}(t)}{l_0} \xi +\dot{d}(t) \right)
|\phi'(\xi,t)|^2 \right]_{-\frac{l_0}{2}}^{\frac{l_0}{2}} ,
\label{exp_val}
\end{eqnarray}
where $\phi'(\xi,t)= \partial_\xi \phi(\xi,t)$.
\textit{Hint:} The energy rate can be computed as 
\begin{equation}
\dot{E}(t)=\frac{\d }{\d t}\braket{\phi}{H(l)
\phi}=\frac{\ii}{\hbar}\big(\braket{\phi }{K H\phi}- \braket{K H
\phi}{\phi }\big)+\braket{\phi }{\dot{H} \phi},
\label{eq:ev_exp_val}
\end{equation}
where $K(l,d)$ is the operator  in~(\ref{eq:transport}).  Pay  attention to the domains.
\end{ex}

\begin{ex}{[Rigid translation]} 
Find the Schr\"odinger operator in the moving reference frame in the case of a translation without dilation of the walls:
\begin{equation}
\dot{l}=0, \qquad d(t)=d_0+vt.
\end{equation}
\end{ex} 

\subsection{An Example: The Accelerating Box}

\begin{figure}[!ht]
   \centering
   \includegraphics[scale=0.35]{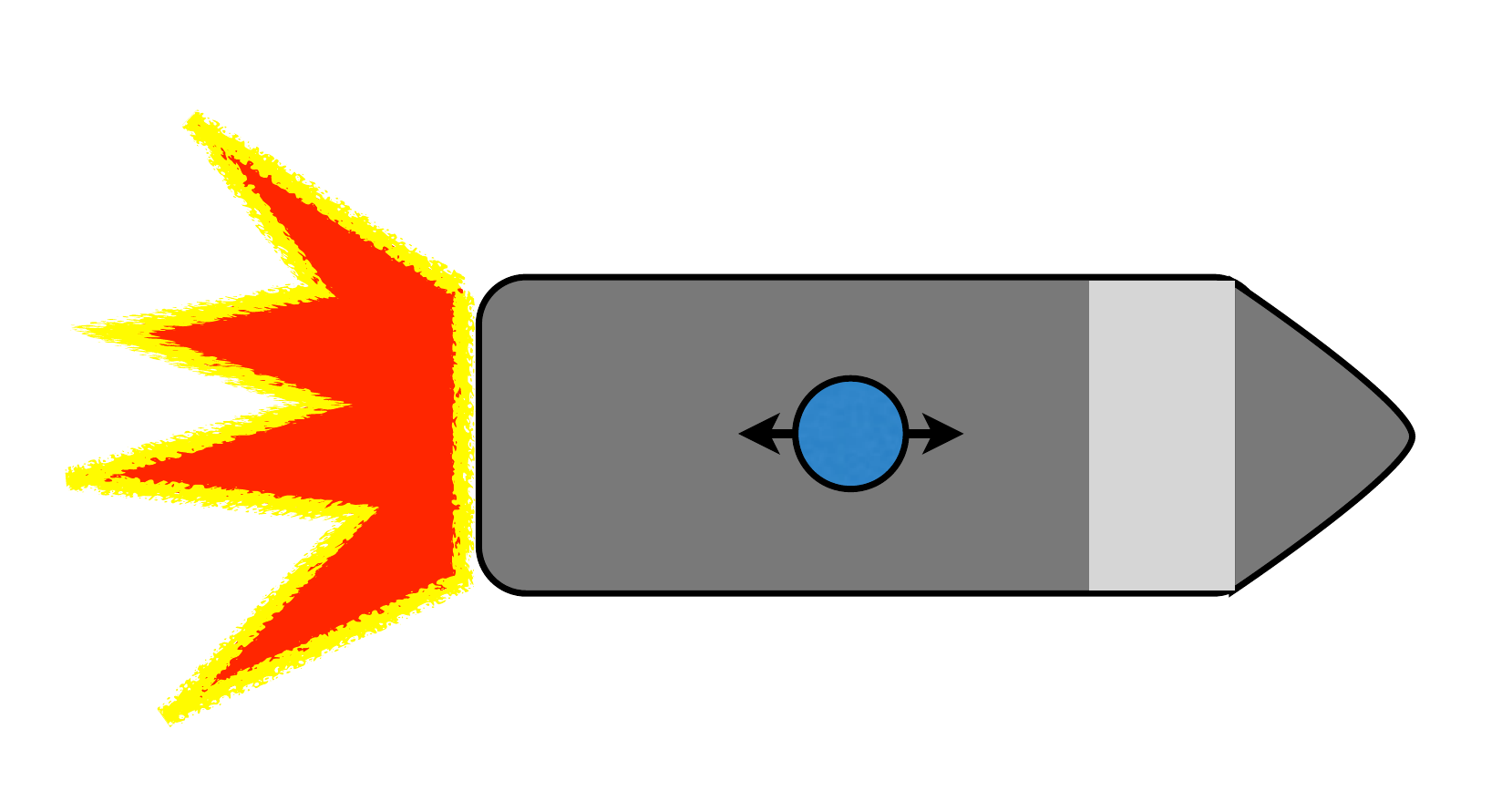}
   \caption{The particle in an accelerating box.}
   \label{fig:elev}
 \end{figure}

Let us consider the case in which the box is moving with a constant acceleration $g$, without dilating: 
\begin{equation}
\dot{l}=0, \qquad d(t)=d_0-\frac{1}{2}gt^2.
\end{equation}
Since the particle is confined in a accelerating rigid box this  is the problem of a particle in a rocket, see Fig.~\ref{fig:elev}.
If we assume $l=l_0$ we find:
\begin{equation}
H=\frac{p^2}{2m}+gtp,
\end{equation}
with Dirichlet boundary conditions.
We can now apply the gauge (unitary) transformation
\begin{equation}
G(t): L^2(I) \rightarrow
L^2(I) : \phi \mapsto \chi =
G(t)\phi\,,
\end{equation}
with 
\begin{equation}
\chi(\xi) = (G(t)\phi)(\xi)=\e^{\frac{\ii}{\hbar}
\left(mg\xi t-\frac{1}{6}mg^2t^3\right)}\phi (\xi)\,,
\end{equation}
in such a way that the Hamiltonian becomes
\begin{equation}
G(t)H G^\dagger(t)=\frac{p^2}{2m}-\frac{1}{2}mg^2t^2.
\end{equation}
Therefore, the Schr\"odinger equation reads
\begin{equation}
\ii \hbar \frac{\d}{\d t} \chi(t)=\left(\frac{p^2}{2m
}- mgx\right)\chi(t),
\end{equation}
describing a particle in a constant gravitational field, in agreement with the equivalence principle. 
\begin{ex}
The eigenfunctions of the operator $\frac{p^2}{2m }-mgx$ belonging to the eigenvalue $E$ are linear combinations of the two Airy functions \cite{Abra}:

\begin{equation}
\phi(x)=c_1 \mbox{Ai}\left(\frac{x}{l}-\frac{E}{mgl}\right)+c_2 \mbox{Bi}\left(\frac{x}{l}-\frac{E}{mgl}\right).
\end{equation}
Prove that the permitted values of energy are the solutions of:
\begin{equation}
 \mbox{Ai}\left(-\frac{1}{2}-\frac{E}{mgl}\right)\mbox{Bi}\left(\frac{1}{2}-\frac{E}{mgl}\right) - \mbox{Ai}\left(\frac{1}{2}-\frac{E}{mgl}\right)\mbox{Bi}\left(-\frac{1}{2}-\frac{E}{mgl}\right)=0,
\label{eq:Airy}
\end{equation}
and in particular that the first four are ($\epsilon=\frac{E}{mgl}$), see Fig.~\ref{fig:Airy}:
\begin{center}
\begin{tabular}{c|c}$\epsilon_1$ & $9.86851$\\
\hline
$\epsilon_2$ & $39.4787$\\
\hline
$\epsilon_3$ & $88.8266$\\
\hline
$\epsilon_4$ & $157.914$\\
\end{tabular}
\end{center}
\begin{figure}[!ht]
   \centering
   \includegraphics[scale=0.50]{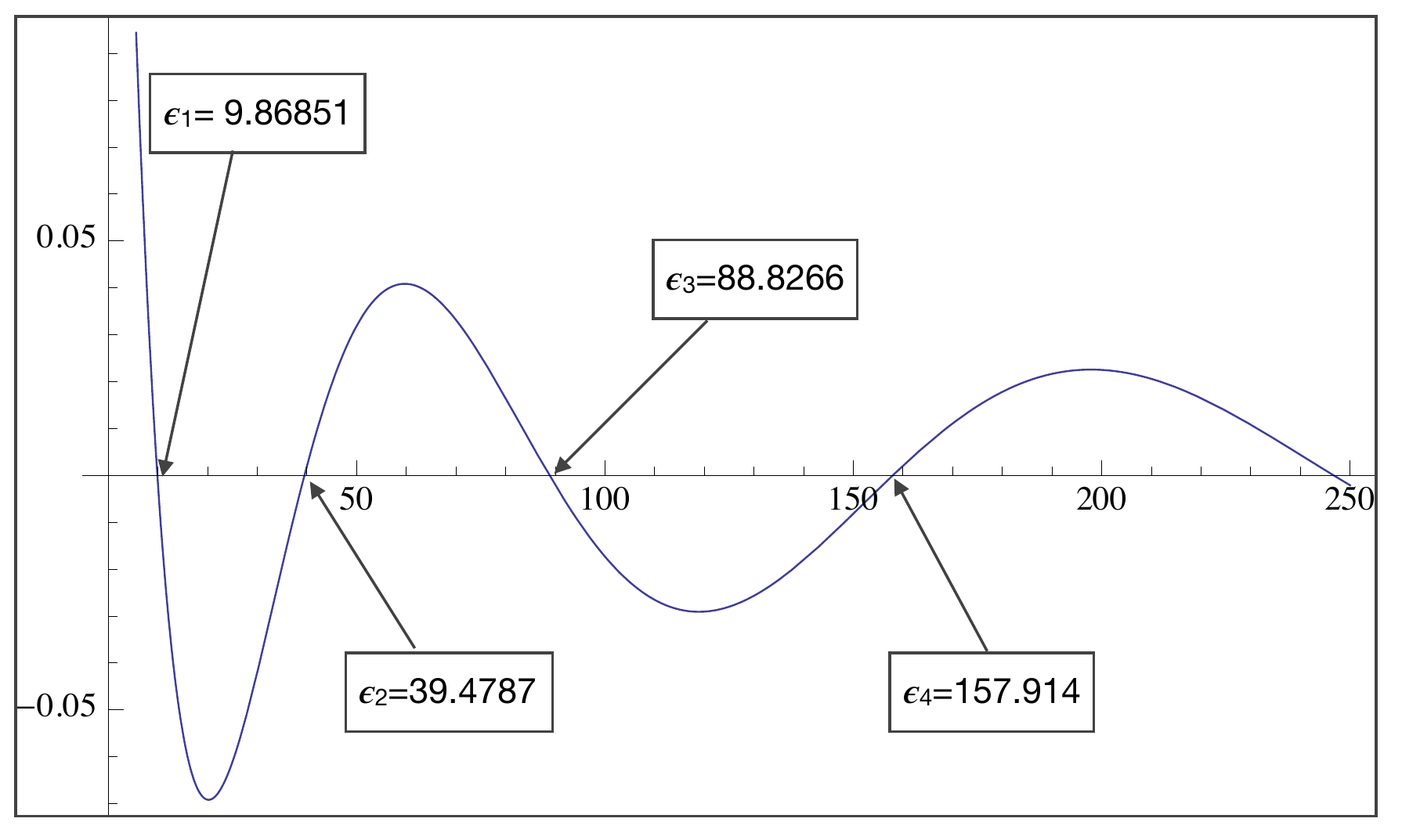}
   \caption{The zeros of the function~(\ref{eq:Airy}) yield the permitted energy levels.}
   \label{fig:Airy}
 \end{figure}
\end{ex}
\section{Lecture 3: Changing boundary conditions}
\label{sec-change}
In this lecture we  will consider a somewhat different situation: we will assume that the walls of the box are fixed, so that $I=[a,b]$ is given, but the way the walls interact with the quantum particle changes in time due to some physical mechanism. That means in our model that the boundary conditions change in time. More generally, the situation we have in mind is that depicted in Fig.~\ref{fig:bend}: a ring with a junction whose properties are time dependent, so that all possible boundary conditions~(\ref{eq:bc2}) can in principle be implemented. The evolution of the particle will be given by a Schr\"odinger equation with a time-dependent Hamiltonian:
\begin{equation}
\ii\hbar \frac{\d}{\d t} \psi(t) = T_{U(t)} \psi(t),
\label{eq:ScUt}
\end{equation}
with $\psi(0)=\psi_0$.
Here $T_{U(t)}= p^2/2m$ on $\mathcal{H}^2(I)\subset L^2(I)$ with boundary conditions~(\ref{eq:bc2}), with $U=U(t)$. Thus the Hamiltonian depends on time through its boundary conditions $U(t)$. 

This model can be implemented in a superconducting quantum interference device (SQUID) with a tunable junction, obtained by replacing the junction with an additional flux loop \cite{Vion,Cosmelli,Mooij}. This can be an experimental realization of a continuous change among different topologies~\cite{Wilczek}.

We will not study this problem in full generality, but instead we will consider the particular case of   boundary conditions rapidly alternating between two  values $U$ and $V$ (for example they periodically alternate between Dirichlet and Neumann): 
\begin{equation}
U(t)=
\begin{cases}
V,& 2k \tau \leq t < (2k+1) \tau, \\
U,& (2k+1) \tau \leq t < (2k+2) \tau,
\end{cases}
\end{equation} 
with $k=0,1,2,\dots$, and $\tau$ the time period. 

In this scenario, since the Hamiltonian is constant during the period among the changes, the Schr\"odinger equation~(\ref{eq:ScUt}) can be explicitly integrated (the trick is revealed: the reason for the funny time-dependence of $U(t)$ is just this!) and at time $2 N \tau$ we end up with
\begin{eqnarray}
\label{eq:2evol}
\psi(2N\tau)&=& \underbrace{\left(\e^{-\ii \tau T_{U}/\hbar} \e^{-\ii \tau T_{V}/\hbar}\right)
\left(\e^{-\ii \tau T_{U}/\hbar} \e^{-\ii \tau T_{V}/\hbar}\right) \dots
\left(\e^{-\ii \tau T_{U}/\hbar} \e^{-\ii \tau T_{V}/\hbar}\right)}_{N \; \mathrm{ times}} \psi_0
 \nonumber \\
&=& \quad \left(\e^{-\ii \tau T_{U}/\hbar} \e^{-\ii \tau T_{V}/\hbar}\right)^N \psi_0.
\end{eqnarray}
We want to study the behavior of the dynamics in the limit $N\to \infty$, when the time interval between the switches $\tau=t/N$ goes to zero, the number of switches goes to infinite, while the total time of the evolution $2 N\tau= 2t$ is kept constant. Therefore, we are led  to the study of the limit of the product formula:
\begin{equation}
\label{eq:evollim?}
\lim_{N\to\infty} \left(\e^{-\ii t T_{U}/N \hbar} \e^{-\ii  t T_{V}/N\hbar}\right)^N .
\end{equation}
\begin{ex}
Prove that, given two matrices $A$ and $B$ the \emph{Lie product formula} holds
\begin{equation}
 \left( \e^{- \ii A/N} \e^{-\ii B/N}\right)^N   \to  \e^{-\ii C},
\label{eq:Trotter}
\end{equation}
with $C=A+B$.

\textit{Hint}: use the telescopic  equation
\begin{equation}
D^N   -  E^N=
 \sum_{k=0}^{N-1} D^k (D-E)  E^{N-1-k},
\end{equation}
with $D= \e^{- \ii A/N} \e^{-\ii B/N}$ and $E= \e^{-\ii (A+B)/N}$, together with the estimates
\begin{equation}
D-E=\e^{- \ii A/N} \e^{-\ii B/N}- \e^{-\ii (A+B)/N} = -\frac{1}{2N^2} [A,B] +O\left(\frac{1}{N^3}\right).
\end{equation}
\end{ex}
Trotter in~\cite{Trotter} proved that Lie's product formula~(\ref{eq:Trotter}) is still valid when applied to (unbounded) self-adjoint  operators $A$ and $B$, such that their sum $C=A+B$ is still (essentially) self-adjoint on the intersection of their domains, $D(C)= D(A)\cap D(B)$.

Well, this seems to be our case: $t T_U/\hbar$ and $t T_V/\hbar$ are unbounded self-adjoint operators, so the limit of the Lie-Trotter product formula~(\ref{eq:evollim?}) should be
$\exp(-\ii (T_U+T_V)t/\hbar)$. Now we get 
\begin{equation}
 T_{U}+T_{V} = \frac{p^2}{2 m} + \frac{p^2}{2 m} = 2 \frac{p^2}{2m},
\end{equation}
on the common domain $D(T_U)\cap D(T_V)$. Therefore,
\begin{equation}
\label{eq:evollim}
\left(\e^{-\ii t T_{U}/\hbar N} \e^{-\ii  t T_{V}/\hbar N}\right)^N \to
\e^{-\ii 2 t T_{W}/\hbar},
\end{equation}
when $N\to\infty$, and we arrive at the intelligible result that rapidly alternating free evolutions with two boundary conditions $U$ and $V$ yield again a free evolution, possibly with different boundary conditions $W$. 

Unfortunately, life is not so easy: Trotter's theorem does not hold, because the intersection of the the domains $D(T_W)=D(T_U)\cap D(T_V)$ is too small, being defined by too many constraints (those of $U$ and those of $V$). For example, let us consider Dirichlet, $U=-\I$, and Neumann, $V=\I$, boundary conditions. In this case $D(T_{-\I})\cap D(T_{\I})=\{\psi\, | \, \Psi=\Psi'=0\}$ and in this domain the kinetic operator $p^2/2m$ is symmetric but not self-adjoint! In other words, our common sense tells us that we should get $p^2/2m$ with some boundary conditions $W$, but we do not know, even heuristically, which ones, because the  boundary conditions $U$ and $V$ fight each other and yield too many constraints!

Math again comes to our aid: The right objects to look at are not the operators $T_U$ and $T_V$, but the \emph{quadratic forms} associated to them, that is their expectation values
$t_U(\psi)=\braket{\psi}{T_U \psi}$ and $t_V(\psi)=\braket{\psi}{T_V \psi}$, measuring the kinetic energy of the particle in state $\psi$. The crucial fact is that the quadratic form $a(\psi) = \braket{\psi}{A \psi}$ associated to an unbounded operator $A$ makes sense on a  domain $D(a)$  which is larger than the domain of $A$, namely $D(A)\subset D(a)$.
Thus the sum of two quadratic forms $a(\psi)+b(\psi)$ is well defined on wave functions $\psi\in D(a)\cap D(b)$ for which the operator sum $A+B$ is not defined, since $\psi\notin D(A)\cap D(B)$.

The extension of  Trotter's theorem to quadratic forms reads as follows \cite{Kato-Trotter,Lapidus}: 
\begin{theorem}{(Lapidus)}
Let  $a$ and $b$ be the quadratic forms associated to $A$ and $B$. If $A$ and $B$ are self-adjoint and bounded below, and $D=D(a)\cap D(b)$ is dense, then Trotter's formula~(\ref{eq:Trotter}) holds with
\begin{equation}
C=A\dot{+}B,
\label{eq:Kato1}
\end{equation}
the unique operator associated to the quadratic form $a+b$. The operator $C$ is called the \emph{form sum} of $A$ and $B$. 
\end{theorem}

 In our case, we will see that the domains of the quadratic forms of the kinetic energy $t_U(\psi)$ involve \emph{only} the boundary values~(\ref{eq:bc21}) of the wave function $\Psi$ and \emph{not} the boundary values of its derivative $\Psi'$. This will imply that the sum of the kinetic energies 
\begin{equation}
t_W(\psi)= \frac{t_U(\psi)+ t_V(\psi)}{2}
\label{eq:sumquad}
\end{equation}
is always the quadratic form associated to a self-adjoint operator $T_W$, namely
\begin{equation}
t_W(\psi)= \braket{\psi}{T_W \psi}, \qquad \psi\in D(T_W).
\end{equation}
For example, in the case of Dirichlet and Neumann we will see that $D(t_{-\I})= \{\psi\, | \, \Psi=0\}$ and $D(t_{\I})= \{\psi\, | \, \text{no bound. conds.}\}$, so that $D(t_{-\I})\cap D(t_{\I})=\{\psi\, | \, \Psi=0\}=D(t_{-\I})$: alternating Dirichlet and Neumann give Dirichlet.

Summing up, we get that the limit~(\ref{eq:evollim}) holds with 
\begin{equation}
T_{W}=\frac{T_U\ \dot{+}\ T_V}{2},
\label{eq:recipe}
\end{equation}
the form sum of $T_U$ and $T_V$.

\subsection{From operators to quadratic forms} 
\label{sec-qforms}

For any $\psi \in \mathcal{H}^2(I)$, one integration by parts yields
\begin{eqnarray}
\braket{ \psi}{ \frac{p^2}{2m} \psi } &=& -\frac{\hbar^2}{2m}\int_a^b \overline{\psi(x)} \psi''(x)\, \d x 
\nonumber\\
&=&  \frac{\hbar^2}{2m}\int_a^b \overline{\psi'(x)} \psi'(x)\, \d x
- \frac{\hbar^2}{2m} \left(\overline{\psi(b)} \psi'(b) - \overline{\psi(a)} \psi'(a)\right)
\nonumber\\
&=& \frac{\hbar^2}{2m}\left( \| \psi' \|^2 - \langle \Psi | \Psi' \rangle \right),
\label{eq:tU0}
\end{eqnarray}
where $\Psi$ and $\Psi'$ are the two-dimensional vectors of boundary data defined in~(\ref{eq:bc21}).
\begin{ex}
Prove that Eq.~(\ref{eq:tU0}) holds.
\end{ex}

Notice that, at variance with our starting point, the last line of~(\ref{eq:tU0}) involves only the first derivative of $\psi$, and thus makes sense on a larger domain. Moreover, by using 
the boundary conditions~(\ref{eq:bc2})-(\ref{eq:bc21}) we will trade the boundary values of the derivative for the boundary values of the function in~(\ref{eq:tU0}), obtaining the following expression for the quadratic form associated to $T_U$:
\begin{equation}
t_U(\psi) = \frac{\hbar^2}{2m}\left( \| \psi' \|^2 + \Gamma_U(\Psi) \right), 
\label{eq:quadraticform}
\end{equation}
with $\Gamma_U(\Psi)$ a quadratic form of the boundary vector $\Psi$.

The form $t_U(\psi)$ is well defined for any integrable function with square integrable \emph{first} (distribution) derivative, 
$\mathcal{H}^1(I)=\{\psi\in L^2(I)\ |\ p\psi\in L^2(I)\}$ (the first Sobolev space). In order to get the explicit expression of $\Gamma_U$ and the precise domain $D(t_U)$ we have to distinguish among three possibilities according to the number of eigenvalues $u_1$ and $u_2$ of $U$ equal to $-1$.
\begin{enumerate}
\item
\label{case:K}
If both the eigenvalues of $U$ are different from $-1$, such as in the case of the Neumann boundary conditions, then $(\I+U)$ is invertible, and the boundary values of the derivative can be expressed in terms of the boundary values of the function:
\begin{equation}
\Psi' =  -\frac{\ii}{l_0} \frac{\I-U}{\I + U} \Psi.
\label{eq:Phi'Phi}
\end{equation}
Therefore we get
\begin{eqnarray}
\Gamma_U (\Psi) = \frac{\ii}{l_0} \langle\Psi | \frac{\I-U}{\I + U} \Psi \rangle,
\label{eq:GammaU1}
\end{eqnarray}
and $D(t_U) = \mathcal{H}^1 (I)$, with no constraints on the boundary values $\Psi$.

\item
\label{case:-1}
If $U$ has only one eigenvalue equal to $-1$, namely $u_1=-1$ and $u_2\neq -1$, then, 
if we call $\xi$ the normalized eigenvector corresponding to the eigenvalue $-1$, and $\xi^\perp$ its orthogonal, from~(\ref{eq:bc2}) we get 
\begin{equation}
\langle \xi | \Psi \rangle = 0, \qquad \langle \xi^\perp| \Psi' \rangle = 
-\frac{\ii}{l_0} \frac{1-u_2}{1+u_2}\,  \langle \xi^\perp| \Psi \rangle .
\label{eq:perp0}
\end{equation}
Therefore,
\begin{equation}
\bra{\xi} \Psi\rangle = 0, \qquad \Gamma_U(\Psi)= \frac{\ii}{l_0} \frac{1-u_2}{1+u_2}\,  |\langle \xi^\perp| \Psi \rangle|^2 .
\label{eq:perp1}
\end{equation}
\begin{ex}
Prove Equations~(\ref{eq:perp0})  and (\ref{eq:perp1}).
\end{ex}

\item
\label{case:Dirichlet}
Finally, if $u_1=u_2=-1$, i.e.\ in the case of the Dirichlet boundary conditions, then $U=-\I$, so that
\begin{equation}
\Psi= 0, \qquad  \Gamma_{-\I}(\Psi) = 0 .
\end{equation}

\end{enumerate}

\subsection{Composition law of boundary conditions}

Now we can use the quadratic form~(\ref{eq:quadraticform}) to evaluate the limit of the alternating dynamics~(\ref{eq:evollim}), according to the recipe~(\ref{eq:recipe}).
By cooking our equations we will  prove that the boundary conditions $W$ are given by the composition law:
\begin{equation}
\label{eq:2evollim}
W= U\star V = V\star U,
\end{equation}
where $\star$ is a commutative and associative product on the boundary unitary operators.

The evaluation of the emergent  boundary condition $W$ in~(\ref{eq:2evollim}) requires the computation of the sum of the kinetic energies~(\ref{eq:sumquad})
 and the evaluation of the domain
\begin{equation}
D(t_W) = D(t_U) \cap D(t_V).
\end{equation} 
Again, we distinguish various cases according to  the number of eigenvalues $-1$:
\begin{enumerate}
\item
When both $U$ and $V$  have no eigenvalues equal to $-1$, we get 
\begin{equation} 
\Gamma_W = (\Gamma_U + \Gamma_V)/2,
\end{equation}
 with no constraints on the wave-function boundary values $\Psi$. By using the expression~(\ref{eq:GammaU1})  we obtain (prove it!)
\begin{equation}
W = U \star V =  \frac{\I-\frac{1}{2}\left(\frac{\I-U}{\I+U} + \frac{\I-V}{\I+V} \right)}
{\I+\frac{1}{2}\left(\frac{\I-U}{\I+U} + \frac{\I-V}{\I+V} \right)} .
\label{eq:regular}
\end{equation}

\item
If $-1$ is a nondegenerate eigenvalue of $U$, and $V$ has no eigenvalues equal to $-1$, then $D(t_W)= D(t_U)$, with the only constraint $\braket{\xi}{\Psi} = 0$. Therefore, the boundary forms $\Gamma_U$ and $\Gamma_V$ are nonzero and add up only on the orthogonal subspace, spanned by $\xi^\perp$. It is easy to see that
\begin{equation}
W= U\star V = - \ket{\xi} \bra{\xi} + w_2 \ket{\xi^\perp} \bra{\xi^\perp},
\label{eq:-1w2}
\end{equation} 
where $w_2$ is a function of $u_2$, $\xi$, and $V$.
\begin{ex}
Derive Eq.~(\ref{eq:-1w2}) and find the explicit expression of $w_2$.
\end{ex}

\item 
If  $-1$ is a nondegenerate eigenvalue of both $U$ and $V$, that is $u_1=v_1 = -1$ and $u_2, v_2 \neq -1$, then there are two possibilities:
\begin{enumerate}
\item 
If the eigenvectors of $U$ and $V$ belonging to $-1$ are parallel,
that is $U$ commutes with $V$, then $D(t_W)=D(t_U)=D(t_V)$. Thus, the only constraint is $\bra{\xi}\Psi\rangle=0$
and $W$ has the previous form~(\ref{eq:-1w2}).
\begin{ex}
Find how $w_2$ particularizes in this case to a function of $u_2$ and~$v_2$.
\end{ex}

\item 
If 
the eigenvectors $\xi$ of $U$ and $\eta$ of $V$ belonging to $-1$ are not parallel, then they span the whole space. The constraints $\braket{\xi}{\Psi}=0$ and $\braket{\eta}{\Psi}=0$ imply  Dirichlet's boundary conditions $\Psi=0$, so that
$D(t_W)=D(t_{-\I})$ and 
\begin{equation}
W= U\star V = -\I .
\label{eq:indipUV}
\end{equation}

\end{enumerate}

\item
Finally, in the case $U=-\I$ (or $V=-\I$) then $D(t_W)= D(t_{-\I})$, so that $\Psi=0$ and
\begin{equation}
W = (-\I) \star U = U \star (-\I) =   -\I.
\label{eq:attractor}
\end{equation}
\end{enumerate}

Summing up, the emerging boundary conditions are given by the composition law~(\ref{eq:2evollim}), where the product $\star$ is given by the Cayley transform~(\ref{eq:regular}) for  $U$ and $V$ with no eigenvalues $-1$ (i.e.\ free ends $\Psi$). On the other hand, eigenspaces with eigenvalues $-1$ are absorbing for the product, that is all constraints on the wave-function boundary values $\Psi$ are inherited by $W$. In particular the Dirichlet boundary conditions $-\I$ play the role of an attractor~(\ref{eq:attractor}), and when 
$U$ and $V$ have independent constraints on $\Psi$, then their composition~(\ref{eq:indipUV}) is  Dirichlet, $\Psi=0$.

\section{Conclusions}
\label{sec-conc}

In these lectures we have seen that sometimes examples that are apparently  very simple may conceal an unexpected rich structure. And in fact they can help us to build and understand general schemes. 
Our goal was to communicate the mathematical and the physical ideas in the most simple setting, without all the unnecessary technical complications that a more realistic model inevitably has. We hope to have hit our goal, at least partially. Moreover, we hope that the references we have provided will stimulate the interest of the reader to go beyond this notes and to get directly in touch with a so active field. For instance, the generalization to higher dimensions is nontrivial and  introduce new savory ingredients, because in such a case the Hilbert space of the boundary is infinite dimensional. But that's  another story.

\section*{Acknowledgments}

We thank Manolo Asorey, Andrzej Kossakowski, Beppe Marmo,  Ninni Messina, Daniele Militello, and Saverio Pascazio for stimulating discussions. We also thank Giuseppe Florio and Giancarlo Garnero for interesting suggestions and for carefully reading the manuscript. P.~F.\ would like to thank the local organizers, A.~P.~Balachandran and Sachin Vaidya, for their kindness in inviting him and for the effort they exerted on the organization of the workshop. This work was partially supported by the Italian National Group of Mathematical Physics (GNFM-INdAM), and by PRIN 2010LLKJBX on ``Collective quantum phenomena: from strongly correlated systems to quantum simulators''.

\end{document}